\documentclass[12pt]{article}
\usepackage{epsf,graphics,psfrag}

\psfrag{5.0}[][][1.8]{$5$}
\psfrag{6.0}[][][1.8]{$6$}
\psfrag{7.0}[][][1.8]{$7$}
\psfrag{8.0}[][][1.8]{$8$}
\psfrag{9.0}[][][1.8]{$9$}
\psfrag{10.0}[][][1.8]{$10$}

\psfrag{stau}[][][1.5]{$\widetilde{\tau}_1$}
\psfrag{stop}[][][1.5]{$\widetilde{t}_1$}
\psfrag{char}[][][1.5]{$\widetilde{\chi}_1^\pm$}
\psfrag{neut}[][][1.5]{$\widetilde{\chi}_1^0$}

\psfrag{+2si}[][][1.5]{$+2\sigma$}
\psfrag{+si}[][][1.5]{$+1\sigma$}
\psfrag{-2si}[][][1.5]{$-2\sigma$}
\psfrag{-si}[][][1.5]{$-1\sigma$}

\psfrag{N=5}[][][1.5]{N=5}
\psfrag{N=6}[][][1.5]{N=6}
\psfrag{N=7}[][][1.5]{N=7}
\psfrag{N=8}[][][1.5]{N=8}
\psfrag{N=9}[][][1.5]{N=9}
\psfrag{N=10}[][][1.5]{N=10}

\psfrag{tanb}[][][2.0]{\raisebox{-10mm}{$\tan\beta$}}
\psfrag{MDM}[][][2.0]{\raisebox{-10mm}{$\Delta a_\mu^{\rm SUSY}$}}
\psfrag{# of messengers}[][][2.0]{\raisebox{-10mm}{$N_{\bf{5}}$}}
\psfrag{masses(GeV)}[][][2.0]{masses(GeV)}
\psfrag{Mmess(GeV)}[][][2.0]{\raisebox{5mm}{$M_{mess}$(GeV)}}

\psfrag{8.0e+00}[][r][1.8]{8.0\hspace{5mm}}
\psfrag{6.0e+00}[][r][1.8]{6.0\hspace{5mm}}
\psfrag{4.0e+00}[][r][1.8]{4.0\hspace{5mm}}
\psfrag{3.0e+00}[][r][1.8]{3.0\hspace{5mm}}
\psfrag{2.0e+00}[][r][1.8]{2.0\hspace{5mm}}
\psfrag{1.0e+00}[][r][1.8]{1.0\hspace{5mm}}
\psfrag{0.0e+00}[][r][1.8]{0\hspace{5mm}}

\psfrag{800.0}[][][1.8]{800}
\psfrag{600.0}[][][1.8]{600}
\psfrag{400.0}[][][1.8]{400}
\psfrag{200.0}[][][1.8]{200}
\psfrag{0.0}[][][1.8]{0}

\psfrag{15.0}[][][1.8]{15}
\psfrag{25.0}[][][1.8]{25}
\psfrag{35.0}[][][1.8]{35}

\psfrag{10-9}[][][1.8]{$\times 10^{-9}$}

\psfrag{tanb=5}[][][1.5]{$\tan\beta=5$}
\psfrag{tanb=30}[][][1.5]{$\tan\beta=30$}

\psfrag{1e+8}[][l][1.8]{$10^8$\hspace{-5mm}}
\psfrag{1e+9}[][l][1.8]{$10^9$\hspace{-5mm}}
\psfrag{1e+10}[][l][1.8]{$10^{10}$\hspace{-7mm}}
\psfrag{1e+11}[][l][1.8]{$10^{11}$\hspace{-7mm}}
\psfrag{1e+12}[][l][1.8]{$10^{12}$\hspace{-7mm}}
\psfrag{1e+13}[][l][1.8]{$10^{13}$\hspace{-7mm}}
\psfrag{1e+14}[][l][1.8]{$10^{14}$\hspace{-7mm}}

\renewcommand{\thefootnote}{\#\arabic{footnote}}
\newcounter{counterA}

\begin{document}

\newcommand{\gtrsim}{ \mathop{}_{\textstyle \sim}^{\textstyle >} }
\newcommand{\lesssim}{ \mathop{}_{\textstyle \sim}^{\textstyle <} }

\renewcommand{\thefootnote}{\fnsymbol{footnote}}
\setcounter{footnote}{0}
\begin{titlepage}

\def\thefootnote{\fnsymbol{footnote}}

\begin{center}

\hfill TU-634\\
\hfill hep-ph/0111155\\
\hfill November, 2001\\
\vskip .5in

{\Large \bf
Anomaly-Mediated Supersymmetry Breaking \\
with Axion
}

\vskip .45in

{\large
Nobutaka Abe, Takeo Moroi and Masahiro Yamaguchi
}

\vskip .45in

{\em
Department of Physics,  Tohoku University\\
Sendai 980-8578, Japan
}

\end{center}

\vskip .4in

\begin{abstract}

We construct hadronic axion models in the framework of the
anomaly-mediated supersymmetry breaking scenario. If the Peccei-Quinn
symmetry breaking is related to the supersymmetry breaking, mass
spectrum of the minimal anomaly-mediated scenario is modified, which
may solve the negative slepton mass problem in the minimal
anomaly-mediated model.  We find several classes of phenomenologically
viable models of axion within the framework of the anomaly mediation
and, in particular, we point out a new mechanism of stabilizing the
axion potential.  In this class of models, the Peccei-Quinn scale is
related to the messenger scale.  We also study phenomenological
aspects of this class of models.  We will see that, in some case, the
lightest particle among the superpartners of the standard-model
particles is stau while the lightest superparticle becomes the axino,
the superpartner of the axion.  With such a unique mass spectrum,
conventional studies of the collider physics and cosmology for
supersymmetric models should be altered.

\end{abstract}
\end{titlepage}

\renewcommand{\thepage}{\arabic{page}}
\setcounter{page}{1}
\renewcommand{\thefootnote}{\#\arabic{footnote}}
\setcounter{footnote}{0}

\section{Introduction}
\label{sec:intro}
\setcounter{equation}{0}

In modern particle physics, symmetries play extremely significant
roles in solving problems related to various fine tunings.  Among
them, in this paper, we consider two serious problems, that is, the
strong CP problem and the gauge hierarchy problem.

Taking account of the instanton effect, the $\Theta$-parameter in QCD,
which is given by\footnote{
The $\Theta$ parameter here is defined in the bases where all the
quark masses are real.}
\begin{eqnarray}
    {\cal L} = \frac{g_3^2}{64\pi^2} \Theta 
    \epsilon^{\mu\nu\rho\sigma} G^a_{\mu\nu}G^a_{\rho\sigma},
\end{eqnarray}
with $g_3$ being the gauge coupling constant for $SU(3)_C$ and
$G^a_{\mu\nu}$ the field strength of the gluon, is constrained to be
smaller than $10^{-9}$ \cite{theta}; otherwise, the electric dipole
moment of the neutron becomes too large to be consistent with the
experimental constraint.  Such a smallness is, however, unnatural in
the standard model, since $\Theta$ is just a parameter in the
Lagrangian.  One smart solution to this problem is to introduce the
Peccei-Quinn (PQ) symmetry \cite{PQ,axion,hadaxion,invaxion}, which is
a spontaneously broken global abelian symmetry which is anomalous
under QCD.  With such a symmetry, which we call $U(1)_{\rm PQ}$,
Nambu-Goldstone boson (called ``axion'' $a$) shows up and couples to
the gluon as
\begin{eqnarray}
    {\cal L} = \frac{g_3^2}{64\pi^2} 
    \left( \Theta + \frac{a}{f_a} \right)
    \epsilon^{\mu\nu\rho\sigma} G^a_{\mu\nu}G^a_{\rho\sigma},
\end{eqnarray}
where $f_a$ is the decay constant of the axion.  In this case, the
$\Theta$ parameter is promoted to a dynamical variable.  Minimizing
the potential of the axion, $\Theta_{\rm eff}\equiv\Theta
+a/f_a\rightarrow 0$, and hence there is no strong CP problem in this
case.  Since the PQ symmetry solves the strong CP problem very
beautifully, it is desirable to implement the PQ symmetry in models of
new physics beyond the standard model.

Another serious fine-tuning in the standard model is related to the
stability of the electroweak scale against radiative corrections; in
the standard model, radiative corrections to the Higgs mass parameter
quadratically diverge and hence the electroweak scale is expected to
be as large as the cutoff scale.  Taking the cutoff at the
gravitational scale, we need more than 30 orders of magnitude of fine
tuning.  

Once the supersymmetry (SUSY) is introduced, quadratic divergences
cancel between bosonic and fermionic diagrams.  Thus, SUSY is
currently regarded as one of the most promising candidate of the
solution to the naturalness problem, and we consider axion model in
supersymmetric framework.\footnote{
For early attempts to construct axion models in the supersymmetric
models, see Refs.\ \cite{SUSYaxion}}
Among various models, in this paper, we consider the anomaly-mediated
SUSY breaking model \cite{AMSB}.  Interestingly, the anomaly-mediated
model may solve the SUSY CP and FCNC problems without fine tuning,
since in this model, all SUSY breaking parameters are well controlled
by the super-Weyl (SW) anomaly.  In addition, this model may provide
solutions to cosmological difficulties such as gravitino problem and
cosmological moduli problem \cite{Cosmo,NPB570-455}.  Thus, we will
study how axion models can be constructed in the framework of the
anomaly-mediated models.  In particular, in some class of the axion
model, we will emphasize that the axion multiplet changes the
prediction of the minimal anomaly-mediated model and that the negative
slepton mass problem may be solved.  We will show that, in this case,
the resulting superparticle mass spectrum is similar to that of the
deflected anomaly mediation \cite{JHEP9905-013}.

Organization of this paper is as follows.  In Section \ref{sec:model},
we introduce several models of axion in the framework of the
anomaly-mediated SUSY breaking.  Then, in Section \ref{sec:pheno}, we
study phenomenology of models introduced in Section \ref{sec:model}.
We summarize our results in Section \ref{sec:summary}.

\section{Model}
\label{sec:model}
\setcounter{equation}{0}

Let us discuss the framework of our model.  Although our mechanism
works in a large class of models with various particle content, we
consider a model with $N_{\bf 5}$-pairs of ${\bf 5}$ and ${\bf
\bar{5}}$ representations in the $SU(5)$ as an example.  The particle
content of our model is given in Table \ref{table:particles}, where
the representations for the standard-model gauge groups as well as the
charge for the $U(1)_{\rm PQ}$ are also shown.  In our model, the
lowest component of $X$ acquires a vacuum expectation value (VEV) and
it breaks the $U(1)_{\rm PQ}$ symmetry.  Thus, we call $X$ as the
axion multiplet.

\begin{table}[t]
\begin{center}
\begin{tabular}{lcccc}
\hline\hline
{} & {$SU(3)_C$} & {$SU(2)_L$} & {$U(1)_Y$} & {$U(1)_{\rm PQ}$} \\
\hline
{$X$} & 
{${\bf 1}$} & {${\bf 1}$} & {$0$} & {$1$} \\
{$Q_i$} & 
{${\bf 3}$} & {${\bf 1}$} & {$-1/3$} & {$-1/2$} \\
{$L_i$} & 
{${\bf 1}$} & {${\bf 2}$} & {$1/2$} & {$-1/2$} \\
{$\bar{Q}_i$} & 
{${\bf \bar{3}}$} & {${\bf 1}$} & {$1/3$} & {$-1/2$} \\
{$\bar{L}_i$} & 
{${\bf 1}$} & {${\bf \bar{2}}$} & {$-1/2$} & {$-1/2$} \\
\hline\hline
\end{tabular}
\caption{Particle content of the model.  The index $i$ is the flavor
index which runs from 1 to $N_{\bf 5}$.}
\label{table:particles}
\end{center}
\end{table}

In addition to the fields in the observable sector, we also introduce
hidden-sector field which is responsible for the SUSY breaking.  We
denote the hidden-sector field as $z$.  In our model, we adopt the
sequestered structure:\footnote{
The function $\xi$ also depends on the observable sector fields such
as quark, lepton, and Higgs multiplets.  Such fields are irrelevant in
studying the axion potential, and we omit these fields in the
expressions.}
\begin{eqnarray}
    K=-3\log \left[ \zeta (z^\dagger ,z) + \xi (X^\dagger , X) \right],
\label{Kahler}
\end{eqnarray}
which may arise, for example, when the hidden and observable sectors
are geometrically separated \cite{sequestered}.

Once the sequestered structure is assumed, phenomenology in the
observable sector is insensitive to how the SUSY is broken in the
hidden sector since the effect of the SUSY breaking is mediated to the
observable sector only by the SW anomaly.  Thus, we only assume that
the SUSY is somehow broken in the hidden sector.  In this framework,
SUSY breakings in the observable sector are only from couplings to the
compensator field $\Phi$, whose VEV is given by
\begin{eqnarray}
    \Phi = 1 + F_\Phi \theta^2.
\end{eqnarray}
Here, $F_\Phi$ is an auxiliary field in the gravitational multiplet,
and its VEV is given by, assuming vanishing cosmological constant,
\begin{eqnarray}
    F_\Phi = \frac{1}{M_*^2}
    \left( W + \frac{1}{3} \frac{\partial K}{\partial z} F_z \right),
\end{eqnarray}
where $M_*\simeq 2.4\times 10^{18}\ {\rm GeV}$ is the reduced Planck
scale.  With this compensator field, relevant part of the supergravity
Lagrangian has the following form \cite{WB}:
\begin{eqnarray}
    {\cal L} &\simeq& 
    \int d^4 \theta \Phi^\dagger \Phi e^{-K/3}
    + \left[ \int d^2 \theta \Phi^3 W + {\rm h.c.} \right]
    \nonumber \\ &\simeq& 
    \int d^4 \theta \Phi^\dagger \Phi 
    \left[ \zeta (z^\dagger,z) + \xi (X^\dagger, X) \right]
    + \left[ \int d^2 \theta \Phi^3 W + {\rm h.c.} \right].
    \label{L_SUGRA}
\end{eqnarray}
Soft SUSY breaking parameters in the observable sector are obtained by
expanding the above Lagrangian in the background with non-vanishing
$F_\Phi$.

Now, let us consider the potential of $X$ and study how the PQ
symmetry can be broken.\footnote{
We use the same expression for the chiral superfield and its bosonic
(lowest) component.}
As we will see below, there are a couple of different approaches, each
of which have different consequences on the mass spectrum of the
squarks and sleptons.  Thus, we will study each of them separately.

Importantly, because of the $U(1)_{\rm PQ}$ symmetry, the function
$\xi$ in the K\"ahler potential depends on the combination $X^\dagger
X$.  {\sl At the tree level}, expanding $\xi$ around $|X|=0$, we
express
\begin{eqnarray}
    \xi_{\rm tree} (X^\dagger , X )
    = M_*^2 \sum_{p=1}^{\infty} c_p \left( \frac{X^\dagger X}{M_*^2} 
    \right)^p
    = c_1 X^\dagger X + \frac{c_2}{M_*^2} (X^\dagger X)^2 + \cdots,
\end{eqnarray}
where $c_p$ are constants.  Notice that, as we will see below,
radiative correction to the function $\xi$ becomes significant in some
case.  In addition, the axion multiplet $X$ is coupled to other fields
in the superpotential as
\begin{eqnarray}
    W_{X{\bf\bar{5}5}} = \lambda_Q X \bar{Q} Q + \lambda_L X \bar{L} L.
    \label{W_X}
\end{eqnarray}
Due to this superpotential, the axion multiplet $X$ couples to the
field strength of the gluon multiplet, and the imaginary part of the
lowest component of $X$ becomes the axion.

At the tree level, and with the simplest K\"ahler potential of
$\xi=c_1 X^\dagger X$, VEV of $X$ is undetermined.  Indeed, after the
rescaling
\begin{eqnarray}
    \hat{X} \equiv \Phi X,~~~
    \hat{Q} \equiv \Phi Q,~~~\cdots,
    \label{hat}
\end{eqnarray}
all the compensators in Eq.\ (\ref{L_SUGRA}) are absorbed in the
``hatted'' superfields.  (Hereafter, ``hatted'' superfields denote
superfields after the rescaling with the compensator $\Phi$, like Eq.\
(\ref{hat}).)  In this case, no potential is generated for $\hat{X}$.
Taking account of the higher dimensional operators and/or radiative
corrections, however, non-trivial potential for $\hat{X}$ is
generated.  In the following subsections, we study several cases.

\subsection{Model 1}

The first approach is to use the radiative correction to the wave
function renormalization factor of $X$ and the higher dimensional
terms in the K\"ahler potential.  Taking account of the wave function
renormalization, at the loop level, we expect the function $\xi$ to become
\begin{eqnarray}
    \xi (X^\dagger , X) = 
    c_1 Z_X(X^\dagger , X) X^\dagger X 
    + \frac{c_2}{M_*^2} (X^\dagger X)^2 + \cdots.
\end{eqnarray}
The $X$-dependence of the wave function renormalization factor $Z_X$
is from the scale dependence; below the scale of the VEV of $X$, all
the particles coupled to $X$ decouple and $Z_X$ does not run.  Thus,
we obtain
\begin{eqnarray}
    Z_X(X^\dagger , X) &=& 
    \sum_n \frac{1}{\Gamma (n+1)}
    \left[ \frac{d^n Z_X}{d(\log\mu)^n} 
    \right]_{\mu=\Lambda}
    \left( \frac{1}{2} \log \frac{X^\dagger X}{\Lambda^2} \right)^n
    \nonumber \\ &=&
    Z_X(\Lambda) 
    + Z'_X(\Lambda) 
    \left( \frac{1}{2} \log \frac{X^\dagger X}{\Lambda^2} \right)
    + \frac{1}{2} Z''_X(\Lambda)  
    \left( \frac{1}{2} \log \frac{X^\dagger X}{\Lambda^2} \right)^2
    + \cdots,
\end{eqnarray}
where $\Lambda$ is an arbitrary scale, and the ``prime'' represents
derivative with respect to $\log\mu$.

Taking account of these effects, the supergravity Lagrangian given in
(\ref{L_SUGRA}) becomes, after the rescaling (\ref{hat}),
\begin{eqnarray}
    {\cal L} &\simeq& 
    \int d^4 \theta Z_X \Bigg\{
    \left[
        1 + \frac{1}{2}\frac{Z'_X}{Z_X} 
        \log\frac{\hat{X}^\dagger \hat{X}}{\Phi^\dagger \Phi}
        + \frac{1}{8}\frac{Z''_X}{Z_X} 
        \left( 
            \log
            \frac{\hat{X}^\dagger \hat{X}}{\Phi^\dagger \Phi} 
        \right)^2
        + \cdots
    \right] \hat{X}^\dagger \hat{X}
    \nonumber \\ &&
    + \frac{c_2}{M_*^2} 
    \frac{(\hat{X}^\dagger \hat{X})^2}{\Phi^\dagger \Phi} 
    + \cdots
    \Bigg\},
    \label{L_X}
\end{eqnarray}
where we performed an $X$-independent rescaling of $X$, and the
potential for $\hat{X}$ is given by
\begin{eqnarray}
    V(X^\dagger , X) &\simeq&
    - \frac{1}{4}
    \left[ \frac{d^2 Z_X}{d(\log\mu)^2} \right]_{\mu=\lambda X} 
    |F_\Phi|^2 \hat{X}^\dagger \hat{X}
    - \frac{c_2}{M_*^2} |F_\Phi|^2 (\hat{X}^\dagger \hat{X})^2 
    + \cdots
    \nonumber \\ &\simeq&
    m_X^2 \hat{X}^\dagger \hat{X} 
    - \frac{c_2}{M_*^2} |F_\Phi|^2 (\hat{X}^\dagger \hat{X})^2 
    + \cdots.
    \label{V(X)}
\end{eqnarray}
With the superpotential given in Eq.\ (\ref{W_X}), the mass for the
$X$ field is given by, at the one-loop level,
\begin{eqnarray}
    m_X^2 (X^\dagger , X) &=& 
    - \frac{1}{4}
    \left[ \frac{d^2 Z_X}{d(\log\mu)^2} \right]_{\mu=\lambda X} 
    |F_\Phi|^2
    \nonumber \\ &=&
    - \left(\frac{1}{16\pi^2}\right)^2 N_{\bf 5}
    \Bigg[ \left( 
        16g_3^2 + \frac{4}{3} g_1^2 \right) \lambda_Q^2
    + \left( 6g_2^2 + 2g_1^2 \right) \lambda_L^2
    \nonumber \\ &&
    - N_{\bf 5} (3\lambda_Q^2 + 2\lambda_L^2)^2
    - 2 (3\lambda_Q^4 + 2\lambda_L^4)  \Bigg] |F_\Phi|^2
    \nonumber \\ &\simeq&
    - \left( \frac{1}{16\pi^2} \right)^2  N_{\bf 5} \left[ 
        16g_3^2 - 5(5N_{\bf 5} +2) \lambda^2 \right] \lambda^2 |F_\Phi|^2,
\end{eqnarray}
where, in the last equality, we used the approximations $g_3\gg g_2$
and $g_1$, and $\lambda\equiv \lambda_Q\sim \lambda_L$.  Hereafter, we
adopt these approximations for simplicity.  Our results are
qualitatively unchanged even if we use the exact formula.

For the spontaneous breaking of the PQ symmetry, $m_X^2$ should become
negative.  This happens when the gauge coupling constants are larger
than the Yukawa couplings $\lambda_Q$ and $\lambda_L$.  Such a
situation can be naturally realized by assuming relatively small
values of the coupling constants $\lambda_Q$ and $\lambda_L$.  Of
course, for a realistic model of SUSY and PQ symmetry breakings,
potential of $\hat{X}$ should be somehow stabilized.  Pomarol and
Rattazzi \cite{JHEP9905-013} proposed to use the inverted hierarchy
mechanism \cite{inverted} to stabilize the potential; if $m_X^2$
changes its sign at some scale with $dm_X^2/d\log\mu >0$, the
potential of $\hat{X}$ has stable minimum at the scale $m_X^2=0$.
Such a scenario is possible when there exists asymptotically-free
gauge interaction with relatively large gauge coupling constant.  We
found, however, that the $SU(3)_C$ gauge interaction cannot play this
role since its gauge coupling constant is not large enough to realize
the inverted hierarchy mechanism. We checked that this is always the
case, by solving renormalization group equations numerically. Pomarol
and Rattazzi suggested to gauge (some part of) the flavor symmetry to
introduce a new gauge interaction for the stabilization.
Unfortunately, it does not work for the axion model since, once $Q$
and $\bar{Q}$ (and/or $L$ and $\bar{L}$) have non-trivial gauge
quantum numbers under extra gauge interaction, the axion potential is
modified and the strong CP problem cannot be solved.

In our model, we do not rely on the inverted hierarchy mechanism,
since there is another simple way to stabilize the potential.  As
suggested in Eq.\ (\ref{V(X)}), the higher dimensional terms in the
K\"ahler potential naturally exist which result in quartic and higher
order terms in the scalar potential.  Assuming the negativity of
$c_2$, the potential given in Eq.\ (\ref{V(X)}) has a minimum, and the
VEV of $\hat{X}$ is given by
\begin{eqnarray}
    f_a^2 \equiv \frac{\langle |\hat{X}|^2 \rangle}{N_{\bf 5}^2}
    = \frac{m_X^2}{2c_2|F_\phi|^2N_{\bf 5}^2} M_*^2
    \simeq \left[ 
        \frac{g_3\lambda}{4\pi^2} \sqrt{\frac{1}{2|c_2|N_{\bf 5}}} M_*
    \right]^2,
\end{eqnarray}
where we assumed $g_3\gg \lambda$.  As one can see, the PQ scale is
suppressed by the loop factor, and it decreases as the coupling
constant $\lambda$ becomes smaller.  Even with $\lambda\sim
O(10^{-1})$, $f_a$ is $O(10^{15}\ {\rm GeV})$, and $f_a$ becomes
$O(10^{12}\ {\rm GeV})$ when $\lambda\sim O(10^{-4})$.
Phenomenological and cosmological implications of this fact will be
discussed later.

Next, let us discuss the mass spectrum of the superparticles in the
MSSM sector.  Since the $X$ field couples to the chiral multiplets
which have standard-model quantum numbers, it affects the soft SUSY
breaking parameters in the observable sector at the loop level.
Indeed, at the scale of the VEV of $X$, the chiral superfields $Q$,
$\bar{Q}$, $L$ and $\bar{L}$ decouple.  Then, the running effects from
the cutoff scale $\Lambda$ to the scale of the VEV of $X$ is via the
combination $\log(X^\dagger X/\Lambda^2) =
\log(\hat{X}^\dagger\hat{X}/\Lambda^2\Phi^\dagger\Phi)$.  Since the
VEV of the highest component of $\hat{X}$ (i.e., the axion multiplet
{\sl after the rescaling}), which is denoted as $F_{\hat{X}}$,
vanishes, we obtain $F_\Phi$ dependence from
$\log(\hat{X}^\dagger\hat{X}/\Lambda^2\Phi^\dagger\Phi)$.  Expanding
this logarithm, we obtain the SUSY breaking parameter at the messenger
scale
\begin{eqnarray}
    M_{\lambda}(M_{\rm mess}) &=& 
    -\frac{b_i-N_{\bf 5}}{4\pi}\alpha_i (M_{\rm mess})F_\Phi,
    \label{M_lam1}\\
    m^2_{\tilde{f}}(M_{\rm mess}) &=& 
    \frac{1}{(4\pi)^2}
    \Bigg[ 2C_i^f\left(b_i-N_{\bf 5}\right)\alpha^2_i(M_{\rm mess})
    \nonumber \\
    && -N_u\alpha_t(M_{\rm mess})
    \Bigg\{
    \frac{13}{15}\alpha_1(M_{\rm mess})+3\alpha_2(M_{\rm mess})+
    \frac{16}{3}\alpha_3(M_{\rm mess})
    \nonumber \\ && 
    -6\alpha_t(M_{\rm mess}) \Bigg\} \Bigg]|F_\Phi|^2,
    \label{m^2_f1}\\
    A_f(M_{\rm mess}) &=& 
    -\frac{y_f(M_{\rm mess})}{4\pi}\sum_{fields \in f}
    \Bigg[2C_i^f\alpha_i(M_{\rm mess})
    -N_u\alpha_t(M_{\rm mess})\Bigg]F_\Phi,
    \label{A_f1}
\end{eqnarray}
where the messenger scale is given by
\begin{eqnarray}
    M_{\rm mess} = \lambda\langle X\rangle,
    \label{M_mess}
\end{eqnarray}
and $\alpha_t$ is top-quark Yukawa coupling, $\alpha_i$ are gauge coupling constants ($i$ runs over the MSSM gauge groups.), $b=(-\frac{33}{5},-1,3)$, and
$C^f$ is the second-order Casimir.  (For the fundamental
representations of $SU(3)_C$ and $SU(2)_L$, $C_3^f=\frac{4}{3}$ and
$C_2^f=\frac{3}{4}$, and for $U(1)_Y$, $C_1^f=\frac{3}{5}Y^2$ with $Y$
being the hypercharge quantum number.)  In addition, $N_u=(1,2,3)$ for
$\widetilde{q}_L^{3rd},\widetilde{t}_R$, and $h_u$ and $N_u=0$ for
other particles.

Contrary to the conventional scenario, there is another important
particle in our model, that is, the axino which is the superpartner of
the axion.  The axino may become the LSP in our model
which has significant implications for collider physics and cosmology.
Axino mass arises from the Lagrangian (\ref{L_X}); expanding the
Lagrangian, the axino mass is given by
\begin{eqnarray}
    m_{\tilde{a}} =
    -\frac{1}{4} \frac{d^2 Z_X}{d(\log\mu)^2} 
    \frac{\langle X^\dagger \rangle}{\langle X\rangle} F_\Phi^\dagger
    \simeq 
    \frac{g_3^2\lambda^2}{16\pi^4} N_{\bf 5} F_\Phi,
    \label{eq:axino-mass}
\end{eqnarray}
where we neglected unimportant phase in the second equality.  As one
can see, the axino mass arises at the two-loop level and is much
smaller than the masses of the superparticles in the MSSM sector.
Thus, in the model 1, the axino becomes the LSP.

Interactions of the axino with the observable sector fields depend on
the PQ charges of the Higgs (and other) fields.  The charge assignment
is model-dependent, and in particular, the charge assignment is related
to how the $\mu_H$- and $B_\mu$-parameters are generated.  This issue
will be discussed later.

\subsection{Model 2}

It is also possible to break the PQ symmetry by higher dimensional
operators in the superpotential.  For this purpose, we introduce extra
chiral multiplet $Y$.  With this superfield, we write down the
following superpotential
\begin{eqnarray}
    W = W_{X{\bf\bar{5}5}} 
    + \frac{1}{\Lambda^{p_X+p_Y-3}} X^{p_X} Y^{p_Y},
\end{eqnarray}
where $p_X$ and $p_Y$ are positive integers, and $\Lambda$ is some
mass parameter which is assumed to be of order the Planck scale.
Obviously, this superpotential is $U(1)_{\rm PQ}$ invariant assigning
the PQ charge $-p_X/p_Y$ to $Y$.

In the background with non-vanishing $F_\Phi$, the scalar potential is
given by
\begin{eqnarray}
    V &=& \left| \frac{1}{\Lambda^{p_X+p_Y-3}} \right|^2
    \left[ p_X^2 \left| \hat{X}^{p_X-1} \hat{Y}^{p_Y} \right|^2 
        + p_Y^2 \left| \hat{X}^{p_X} \hat{Y}^{p_Y-1} \right|^2 \right]
    \nonumber \\ && 
    + (p_X+p_Y-3) \left[ \frac{F_\Phi}{\Lambda^{p_X+p_Y-3}} 
        \hat{X}^{p_X} \hat{Y}^{p_Y}
        + {\rm h.c.} \right].
    \nonumber \\
\end{eqnarray}
Minimizing this potential, we obtain
\begin{eqnarray}
    p_Y |\langle\hat{X}\rangle |^2 =
    p_X |\langle\hat{Y}\rangle |^2 =
    C \left| F_\Phi \Lambda^{p_X+p_Y-3} \right|^{2/(p_X+p_Y-2)},
\end{eqnarray}
where 
\begin{eqnarray}
    C = 
    \left( \frac{p_X+p_Y-3}{p_X+p_Y-1} \right)^{2/(p_X+p_Y-2)}
    p_X^{(p_Y-2)/(p_X+p_Y-2)} p_Y^{(p_X-2)/(p_X+p_Y-2)},
\end{eqnarray}
which is a constant of $O(1)$.  Thus, we obtain
\begin{eqnarray}
    f_a \sim 
    \frac{|\langle\hat{X}\rangle |}{N_{\bf 5}} \sim
    \frac{|\langle\hat{Y}\rangle |}{N_{\bf 5}} \sim
    \frac{\left| F_\Phi \Lambda^{p-3} \right|^{1/(p-2)}}{N_{\bf 5}},
\end{eqnarray}
and
\begin{eqnarray}
    \frac{F_{\hat{X}}}{\hat{X}} = -\frac{p-3}{p-1} F_\Phi,
    \label{F/X(2)}
\end{eqnarray}
where $p=p_X+p_Y$.  In this case, the PQ scale is determined by the
relative size of $F_\Phi$ and $\Lambda$.  Taking $\Lambda\sim M_*$,
for example, and $F_\Phi\sim 100\ {\rm TeV}$, we obtain $f_a\sim
O(10^{10}\ {\rm GeV})$, $O(10^{13}\ {\rm GeV})$, and $O(10^{14}\
{\rm GeV})$, for $p=4$, $5$, and $6$, respectively. Especially, in the 
case of $p=4$, $N_{\bf 5}\gtrsim 10$.

With the VEV given in Eq.\ (\ref{F/X(2)}), we obtain
\begin{eqnarray}
    M_{\lambda}(M_{\rm mess}) &=& 
    -\frac{b_i-N_{\bf 5}}{4\pi}\alpha_i (M_{\rm mess}) F_\Phi,
    \\
    m^2_{\tilde{f}}(M_{\rm mess}) &=& 
    \frac{1}{(4\pi)^2}
    \Bigg[ 2C_i^f \left(b_i-\frac{4(p-2)}{(p-1)^2}N_{\bf 5}\right)
    \alpha_i^2(M_{\rm mess})
    \nonumber \\ && 
    -N_u\alpha_t(M_{\rm mess})
    \Bigg\{ \frac{13}{15}\alpha_1(M_{\rm mess})+3\alpha_2(M_{\rm mess})
    +\frac{16}{3}\alpha_3(M_{\rm mess})
    \nonumber \\ &&
    -6\alpha_t(M_{\rm mess}) \Bigg\} \Bigg] |F_\Phi|^2,
    \\
    A_f(M_{\rm mess}) &=& 
    -\frac{y_f(M_{\rm mess})}{4\pi}
    \sum_{fields \in f} 
    \Bigg[2C_i^f\alpha_i(M_{\rm mess})-N_u\alpha_t(M_{\rm mess})\Bigg]
    F_\Phi.
\end{eqnarray}

One important difference between this model and the model 1 is
the axino mass.  In this model, axino is a linear combination of the
fermionic component in the chiral superfields $X$ and $Y$.  The mass
matrix for these fermions are, after relevant phase rotation to remove
unwanted phases,
\begin{eqnarray}
    {\bf M}_{\rm axino} =
    C^{(p_X+p_Y)/2} p_X^{(2-p_Y)/2} p_Y^{(2-p_X)/2}
    \left( \begin{array}{cc}
            p_X - 1 & \sqrt{p_X p_Y} \\
            \sqrt{p_X p_Y} & p_Y - 1
        \end{array} \right) 
    |F_\Phi|.
\end{eqnarray}
As one can easily see, both fermions acquire masses as large as
$|F_\Phi|$, which is of order $10-100\ {\rm TeV}$.  Thus, in this
model, axino mass is much heavier than the masses of the
superparticles in the MSSM sector and hence the axino cannot be the
LSP. 

\subsection{Model 3}

In the previous models, the PQ symmetry is unbroken in the
supersymmetric limit.  In some case, however, the PQ symmetry can be
spontaneously broken even if the supersymmetry is preserved.  Let us
finally consider such a case.

The simplest superpotential realizing such a situation is
\begin{eqnarray}
    W = W_{X{\bf\bar{5}5}} 
    + \kappa Y' (\bar{X} X - N_{\bf 5}^2 f_a^2),
\end{eqnarray}
where $f_a$ here is some constant, and $Y'$, whose $U(1)_{\rm PQ}$
charge is 0, is a chiral superfield which is singlet under $SU(5)$.
Contrary to the previous models where the PQ scale $f_a$ is somehow
related to the SUSY breaking parameter $F_\Phi$, $f_a$ is an input
parameter in this case.  As one can see, the PQ symmetry is broken
solving the condition for the vanishing $F$-component of $Y'$:
$\partial W/\partial Y'=0$.

In this model, in fact, the PQ symmetry is broken but the masses of
the squarks and sleptons are not modified.  This can be easily seen by
solving the equations of motion.  After the rescalings $\hat{X}=\Phi
X$ and so on, we obtain $\langle\hat{\bar{X}}\hat{X}\rangle
=f_a^2\Phi^2$, by solving $\partial W/\partial \hat{Y}'=0$.  At this
stage, the relative size of $\hat{X}$ and $\hat{\bar{X}}$ is
undetermined since it corresponds to a flat direction in the
supersymmetric limit.  The relative size is determined taking account
the effect of the SUSY breaking.  With non-vanishing $F_\Phi$,
$\hat{Y}'$ acquires a VEV as
\begin{eqnarray}
    \kappa \langle \hat{Y}' \rangle \simeq F_\Phi,
\end{eqnarray}
which gives equal masses to $\hat{X}$ and $\hat{\bar{X}}$.  Then, the
VEVs of these chiral superfields become the same and
\begin{eqnarray}
    \langle \hat{\bar{X}} \rangle = \langle \hat{X} \rangle 
    = N_{\bf 5} f_a \Phi,
\end{eqnarray}
up to an irrelevant phase.  Thus, substituting this VEV into
$\log(\hat{X}^\dagger\hat{X}/\Lambda^2\Phi^\dagger\Phi)$, effect of
the SUSY breaking disappears, and hence the axion multiplet does not
modify the masses of the superparticles.  In this case, extra
mechanism should be introduced to solve the negative slepton mass
problem, and we do not pursue this direction anymore.

\subsection{The $\mu_H$- and $B_\mu$-Parameters}

Before closing this section, let us comment on the $\mu_H$- and
$B_\mu$-parameters.  The $\mu_H$- and $B_\mu$-parameters can be
generated by slightly modifying the model by Pomarol and Rattazzi
\cite{JHEP9905-013}.  The new superpotential we introduce is
\begin{eqnarray}
    W = \lambda_{STX} STX + \lambda_{SST} S^2 T
    + \lambda_{THH} T H_u H_d,
\end{eqnarray}
where $S$ and $T$ are chiral superfield which are singlet under the
$SU(5)$, whose PQ charges are $+1$ and $-2$, respectively.  Notice
that we also assigned the PQ charge $-1$ for the up- and down-type
Higgses. 

As in the case of the model by Pomarol and Rattazzi, kinetic mixing
between $X$ and $S$ appears at the one-loop level:
\begin{eqnarray}
    {\cal L}_{\rm mix} = \int d^4\theta \Phi^\dagger\Phi
    Z_{SX} (\mu = |\lambda X|) S X^\dagger + {\rm h.c.}
    \label{kin-mix}
\end{eqnarray}
In addition, once $X$ acquires a VEV, $S$ and $T$ becomes massive and
these fields can be integrated out.  In particular, by solving the
condition $\partial W/\partial T=0$, we obtain
\begin{eqnarray}
    S = - \frac{\lambda_{THH}}{\lambda_{STX}}
    \frac{H_u H_d}{X},
    \label{<S>}
\end{eqnarray}
and substituting Eq.\ (\ref{<S>}) into Eq.\ (\ref{kin-mix}), we obtain
\begin{eqnarray}
    {\cal L}_{\rm mix} = - \frac{\lambda_{THH}}{\lambda_{STX}}
    \int d^4\theta
    Z_{SX} (\mu = |\lambda \hat{X}/\Phi|)
    \frac{\hat{X}^\dagger}{\hat{X}} \hat{H}_u \hat{H}_d + {\rm h.c.}
\end{eqnarray}
In Model 1 where $F_{\hat{X}}=0$, expanding the above Lagrangian, we
obtain the $\mu_H$- and $B_\mu$-parameters to be
\begin{eqnarray}
    \mu_H &=& - \frac{\lambda_{THH}}{2\lambda_{STX}}
    \left[ \frac{d Z_{SX}}{d \log\mu} \right]_{\mu = \lambda X}
    F_\Phi^\dagger,
    \\
    B_\mu &=&  - \frac{\lambda_{THH}}{4\lambda_{STX}}
    \left[ \frac{d^2 Z_{SX}}{(d \log\mu)^2} \right]_{\mu = \lambda X}
    |F_\Phi|^2.
\end{eqnarray}

\section{Phenomenology}
\label{sec:pheno}
\setcounter{equation}{0}

\begin{table}[t]
\begin{center}
\begin{tabular}{c@{\hspace{3cm}}c}
The model 1 & The model 2 ($p=4$ case)\\
\begin{tabular}[t]{|c|c||c|c|}\hline
$\widetilde{e}_1$ & 144 & $\widetilde{e}_2$ & 265\\\hline
$\widetilde{\tau}_1$ & 104 & $\widetilde{\tau}_2$ & 270\\\hline
$\widetilde{\nu}_e$ & 253 & $\widetilde{\nu}_\tau$ & 248\\\hline
$\widetilde{d}_1$ & 686 & $\widetilde{d}_2$ & 732\\\hline
$\widetilde{b}_1$ & 648 & $\widetilde{b}_2$ & 695\\\hline
$\widetilde{u}_1$ & 690 & $\widetilde{u}_2$ & 728\\\hline
$\widetilde{t}_1$ & 579 & $\widetilde{t}_2$ & 737\\\hline
$\widetilde{\chi}_1^\pm $ & 306 & $\widetilde{\chi}_2^\pm $ & 615\\\hline
$\widetilde{\chi}_1^0 $ & 299 & $\widetilde{\chi}_2^0 $ & 319\\\hline
$\widetilde{\chi}_3^0 $ & 528 & $\widetilde{\chi}_4^0 $ & 616\\\hline
$h^0$ & 117 & $H^0$ & 332\\\hline
$A^0$ & 333 & $H^\pm$ & 345\\\hline
$\widetilde{g}$ & 941 & $-$ & -\\\hline
\end{tabular}
&
\begin{tabular}[t]{|c|c||c|c|}\hline
$\widetilde{e}_1$ & 167 & $\widetilde{e}_2$ & 460\\\hline
$\widetilde{\tau}_1$ & 104 & $\widetilde{\tau}_2$ & 462\\\hline
$\widetilde{\nu}_e$ & 453 & $\widetilde{\nu}_\tau$ & 447\\\hline
$\widetilde{d}_1$ & 1675 & $\widetilde{d}_2$ & 1735\\\hline
$\widetilde{b}_1$ & 1623 & $\widetilde{b}_2$ & 1664\\\hline
$\widetilde{u}_1$ & 1677 & $\widetilde{u}_2$ & 1733\\\hline
$\widetilde{t}_1$ & 1512 & $\widetilde{t}_2$ & 1691\\\hline
$\widetilde{\chi}_1^\pm $ & 706 & $\widetilde{\chi}_2^\pm $ & 1458\\\hline
$\widetilde{\chi}_1^0 $ & 702 & $\widetilde{\chi}_2^0 $ & 711\\\hline
$\widetilde{\chi}_3^0 $ & 1192 & $\widetilde{\chi}_4^0 $ & 1458\\\hline
$h^0$ & 130 & $H^0$ & 712\\\hline
$A^0$ & 716 & $H^\pm$ & 721\\\hline
$\widetilde{g}$ & 2538 & $-$ & -\\\hline
\end{tabular}
\end{tabular}
\end{center}
\caption{Mass spectra of the model 1 and 2 in units of GeV. 
We take $\tan\beta=30$, $M_{mess}=10^{12}\textrm{GeV}$ and
$10^{11}\textrm{GeV}$, $N_{\bf 5}=7$ and 13 for the model 1 and 2,
respectively.  The Wino mass $M_2$ is taken to be 600 GeV for the
model 1 and 1450 GeV for the model 2.}
\label{table:spectrum}
\end{table}

Here we would like to discuss phenomenological issues of the models we
presented. As we mentioned earlier, our models have superparticle
masses identical to those in the deflected anomaly mediation. Some of
our results were already discussed in Refs.~\cite{JHEP9905-013,RSW}.

\subsection{Mass Spectrum}
Let us first consider the model 1. In this model, the soft masses are
given in Eqs.\ (\ref{M_lam1}), (\ref{m^2_f1}) and (\ref{A_f1}), which
are parameterized by the scale of the SUSY breaking $F_{\Phi}$, the
messenger scale $M_{\rm mess}$ and the number of the messenger fields
$N_5$. In addition, we have the $\mu_H$ and $B_{\mu}$ parameters. Here
we do not take any particular mechanism to generate them. $\mu_H$ is
solely determined so as to reproduce the correct electroweak scale,
and $B_{\mu}$ is related to $\tan \beta$, the ratio of the vacuum
expectation values of the two Higgs in the MSSM.  Thus the
superparticle mass spectrum is specified by $F_{\Phi}$, $M_{\rm
mess}$, $N_5$ and $\tan\beta$.

\begin{figure}
    \begin{center}
        \scalebox{0.5}[0.5]{\includegraphics{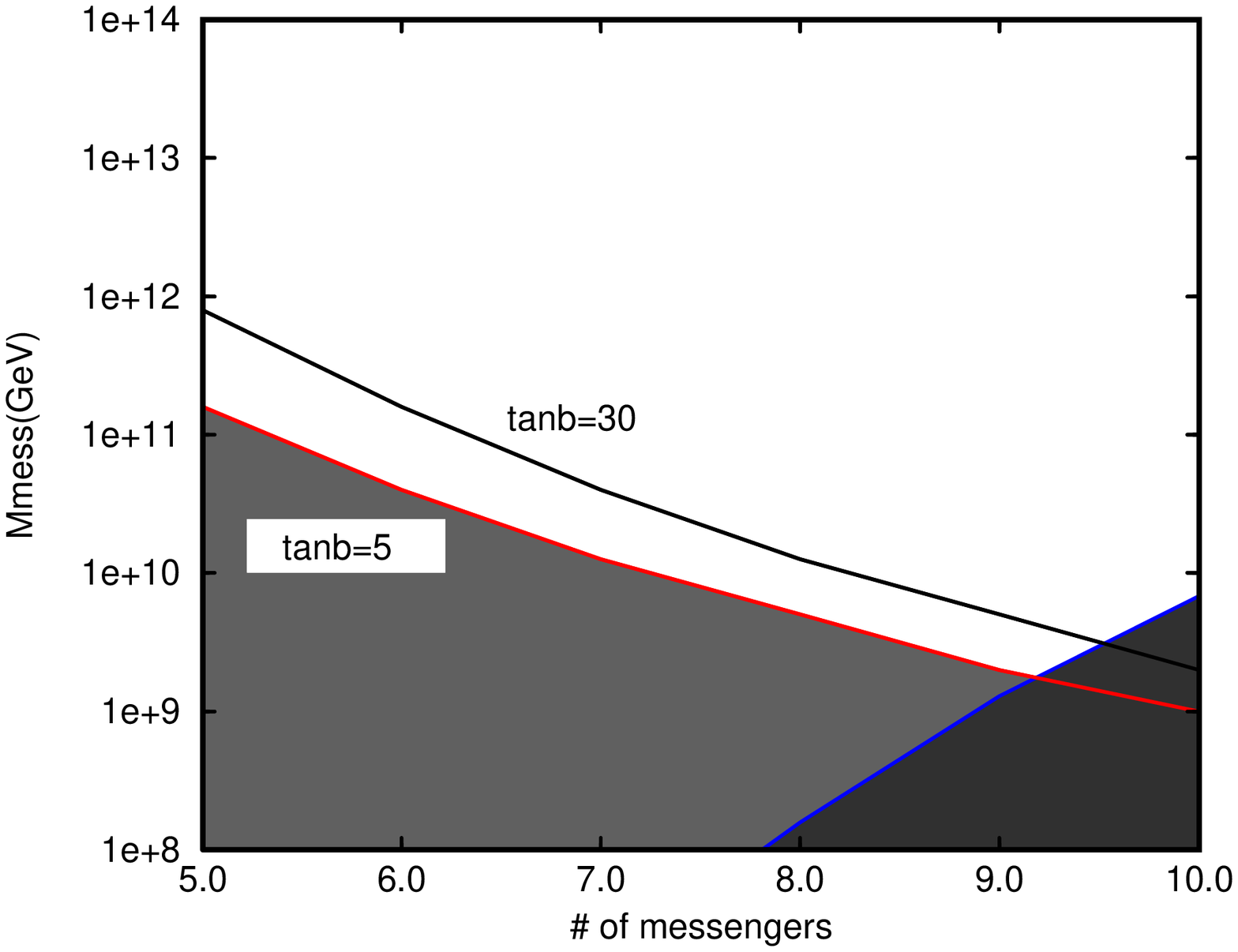}}
        \caption{Lower bound on the messenger scale $M_{\rm mess}$ as 
        a function of the number of messengers $N_{5}$. The
        light-shaded region is excluded by tachyonic slepton for
        $\tan\beta=5$.  The lower limit for $\tan\beta=30$ is also
        shown as a solid line. The dark-shaded region is excluded
        because the gauge coupling constants blow up below the GUT
        scale.}
        \label{fig:map}
    \end{center}
\end{figure}

\begin{figure}
    \begin{center}
        \scalebox{0.5}[0.5]{\includegraphics{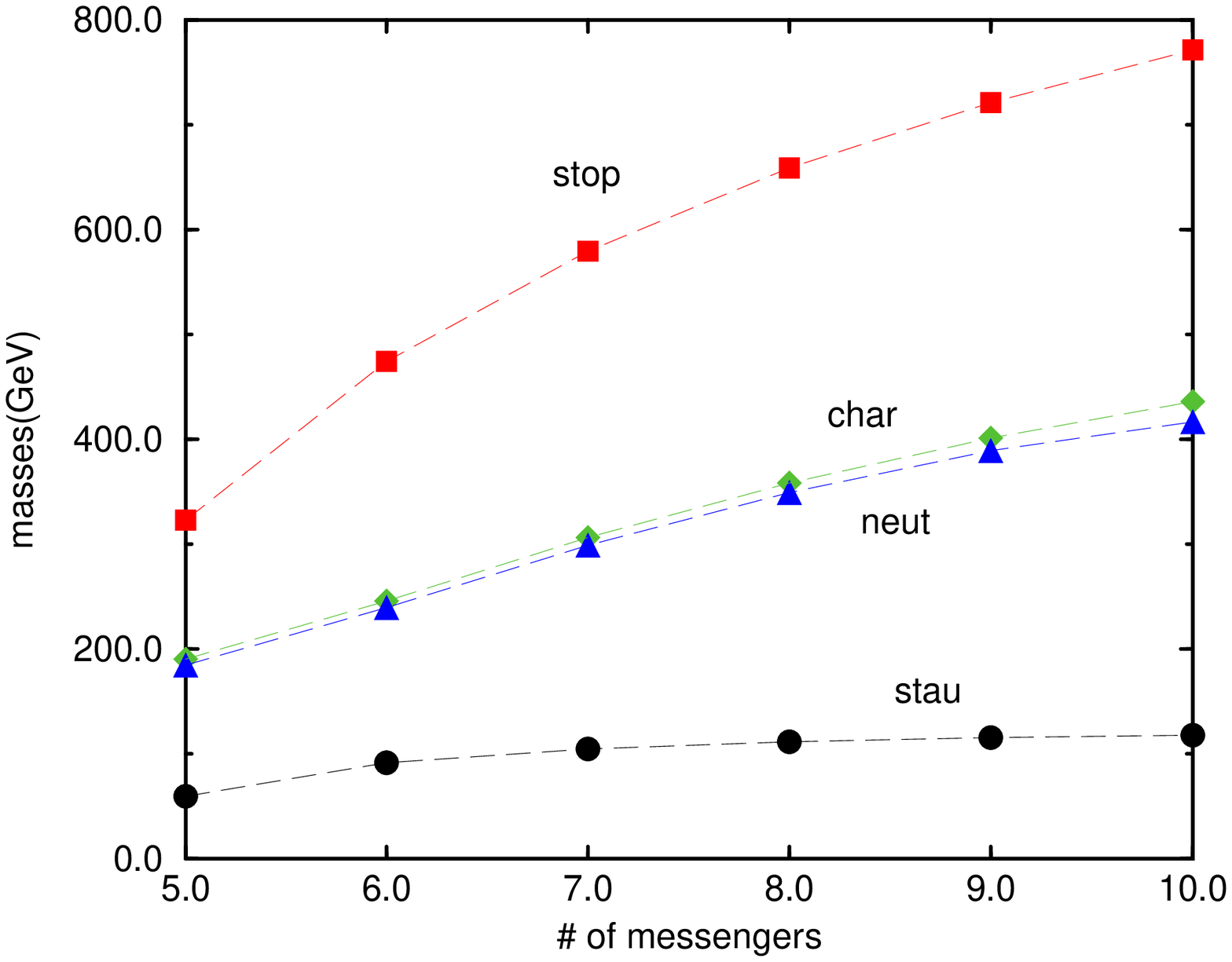}}
        \caption{Mass spectrum of $\widetilde{\tau}_1$,
        $\widetilde{\chi}_1^0$, $\widetilde{\chi}_1^\pm$ and
        $\widetilde{t}_1$, with $\tan\beta=30$, $M_{\rm mess}=10^{12}$
        GeV and $M_2=600$ GeV.}
        \label{fig:spect12}
        \vskip 1cm
        \scalebox{0.5}[0.5]{\includegraphics{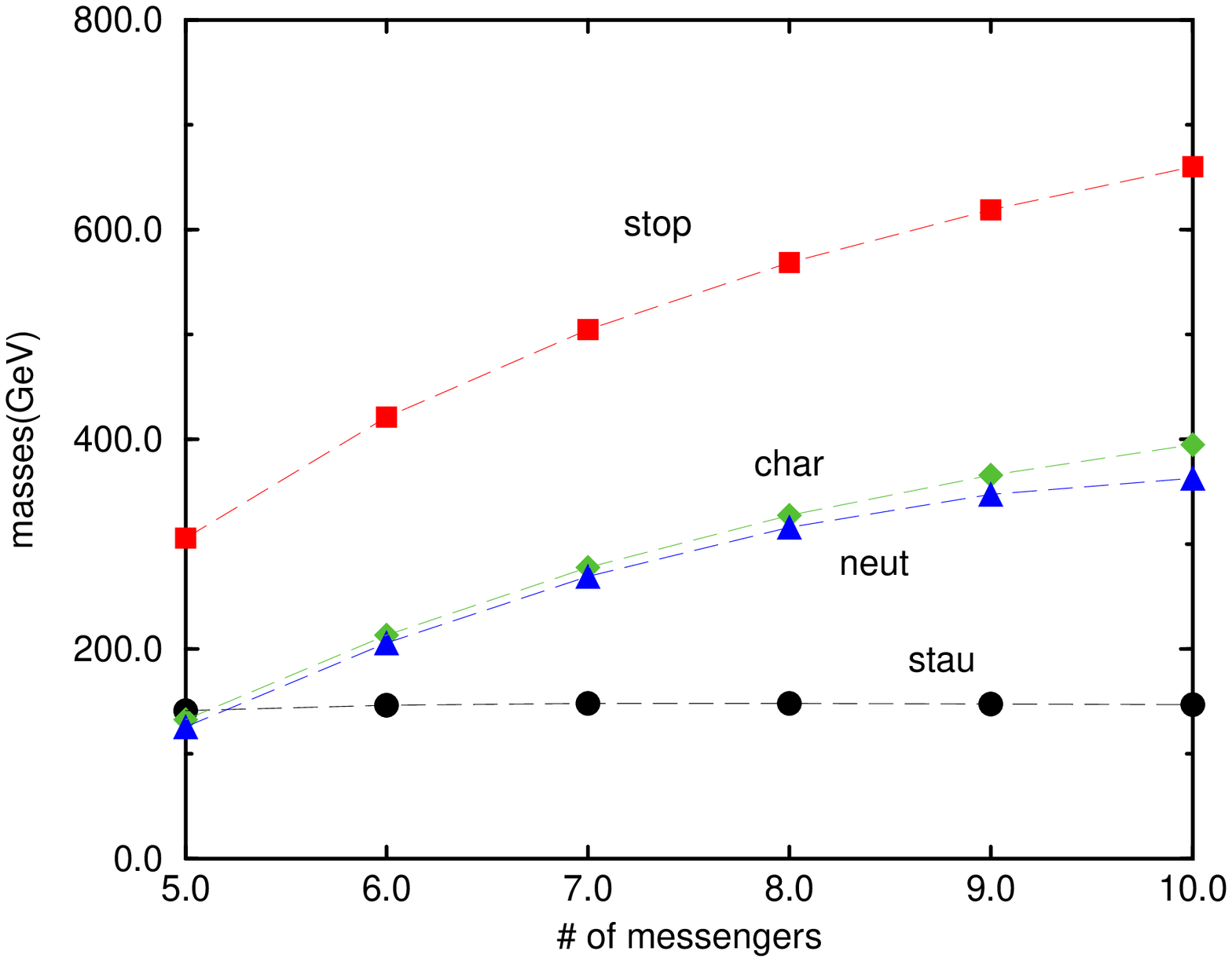}}
        \caption{Same as Fig.\ \ref{fig:spect12}, except
        $M_{\rm mess}=10^{14}$ GeV and $M_2=500$ GeV.}
        \label{fig:spect14}
    \end{center}
\end{figure}

\begin{figure}
    \begin{center}
        \scalebox{0.5}[0.5]{\includegraphics{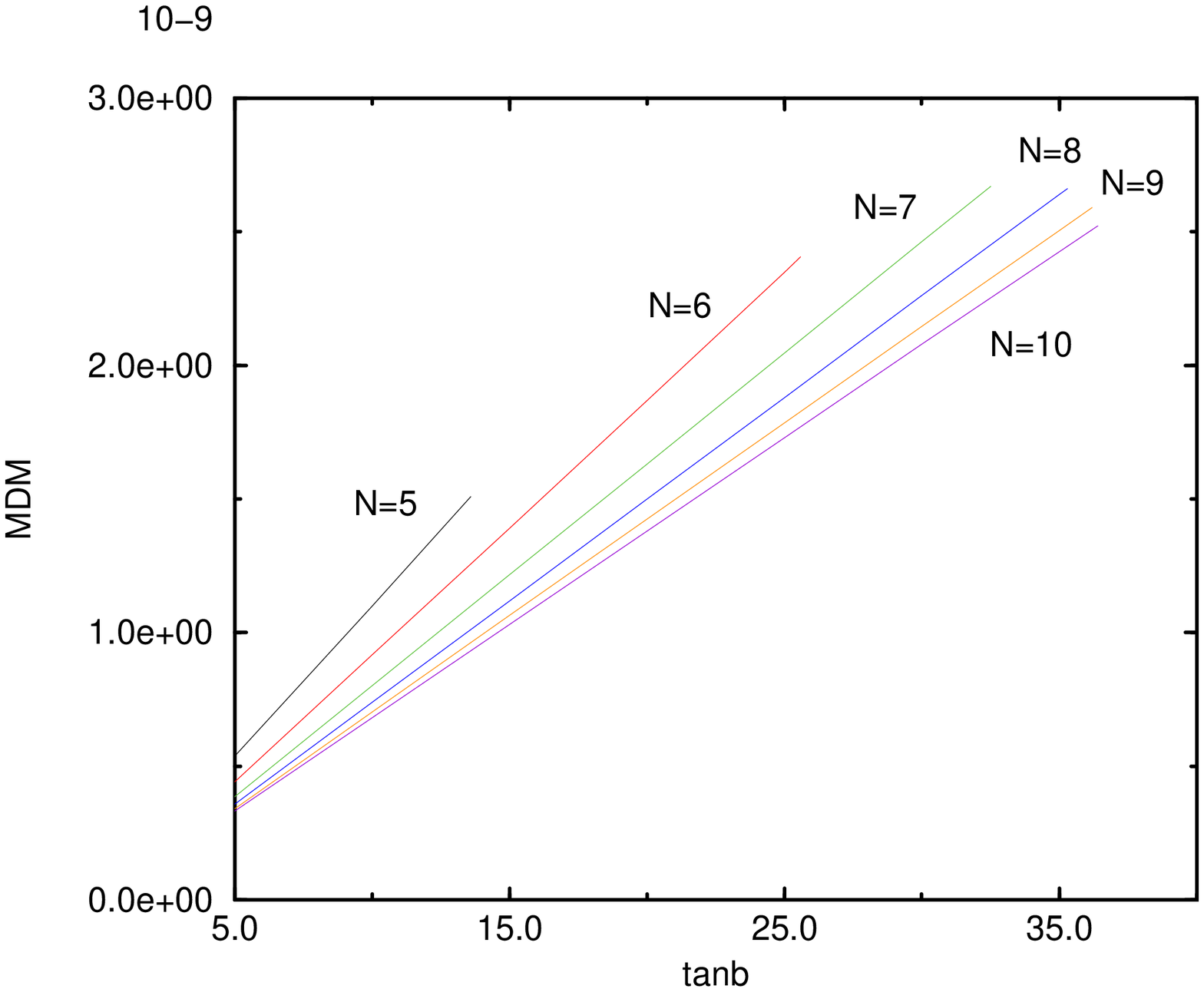}}
        \caption{The SUSY contribution of the muon magnetic dipole 
        moment $a_\mu=\frac{1}{2}(g_\mu -2)$, for $M_{\rm
        mess}=10^{12}$ GeV and $M_2=600$ GeV.}
        \label{fig:mdm12}
        \vskip 1cm
        \scalebox{0.5}[0.5]{\includegraphics{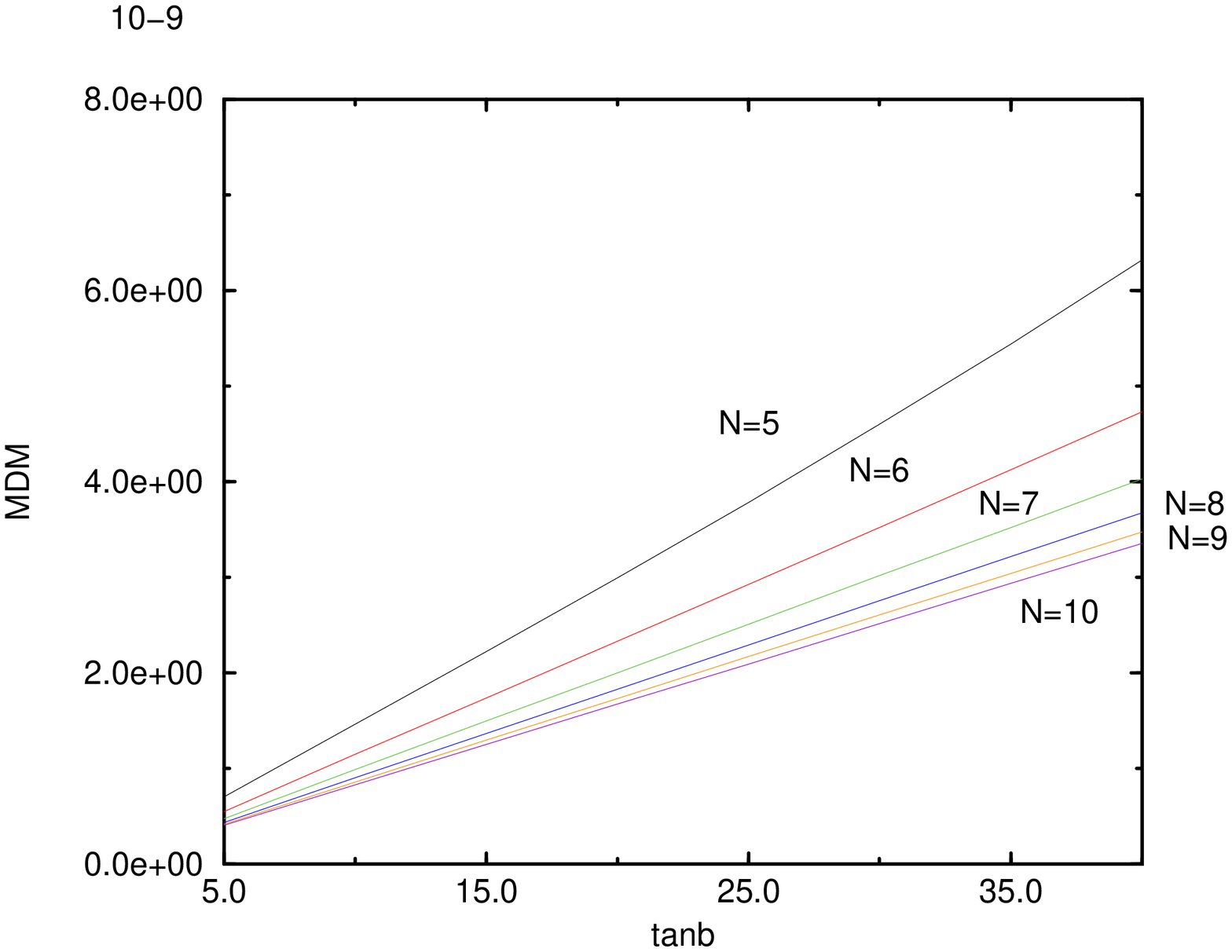}}
        \caption{Same as Fig.\ \ref{fig:mdm12}, except $M_{\rm
        mess}=10^{14}$ GeV and $M_2=500$ GeV.}
        \label{fig:mdm14}
    \end{center}
\end{figure}

To obtain a realistic mass spectrum (positive slepton masses etc.),
the number of the messenger should be large enough.  In fact for
$N_5=3$, the gluino mass vanishes, resulting in an unrealistic mass
spectrum.  For $N_5=4$, we found that the sleptons are too light to
survive the experimental mass bounds, unless the SUSY breaking scale
is very high. This is disfavored, because the other superparticle
masses will exceed 1 TeV, causing the fine-tuning problem.  Thus we
will consider the case $N_5 \geq 5$.  In Fig.\ \ref{fig:map}, we show
the lower bound on the messenger scale $M_{\rm mess}$ as a function of
$N_5$. The bound comes from the two requirements: (\roman{counterA})
the positive slepton mass squared
(\stepcounter{counterA}\roman{counterA}) the perturbativity of the
gauge coupling constants up to the GUT scale.  The former requirement
is severer for smaller $N_5$, while the latter is severer for larger
$N_{5}$. One finds that the messenger scale $M_{\rm mess}$ must be
larger than $\sim 10^9$ GeV.\footnote
{In Fig.\ \ref{fig:map}, we require that all the soft SUSY breaking
mass squared parameters for sleptons be positive to make the figure
being independent of $F_\Phi$.  Notice that such a constraint is
slightly different from the actual experimental constraint based on
the mass eigenvalues.  (See Figs.\ref{fig:spect12} and \ref{fig:spect14}.)
Constraint on $M_{\rm mess}$ is, however, almost unchanged.}

In Figs.\ \ref{fig:spect12} and \ref{fig:spect14}, the masses of some
superparticles are shown for $M_{\rm mess}=10^{12}$ and $10^{14}\ {\rm
GeV}$, respectively.  One finds that the lightest superparticle among
the MSSM particles is almost always a stau, the superpartner of the
tau lepton. Note that in the model 1, the stau is not stable since the
axino is the lightest superparticle and the stau decays to the
axino. We will discuss phenomenological and cosmological implications
to the stau decay.  In addition, it is also notable that the slepton
masses are more enhanced relative to the gauginos if we adopt larger
value of $M_{\rm mess}$.

In Table \ref{table:spectrum}, we show the superparticle mass spectrum
for some typical cases.\footnote
{The Higgs masses are computed by using the effective potential at
one loop level. Higher order corrections will somewhat reduce the
$h^0$ mass.}  
As we mentioned, the lightest superparticle in the MSSM sector is the
stau.  The spectrum is rather compact, and squarks and gluinos are
much lighter for the same stau mass than in the case of gauge
mediation \cite{RSW}.  Another interesting point is that the lightest
neutralino is higgsino-like, which is not achieved in many scenarios
of supersymmetry breaking, including the minimal supergravity model
with all the SUSY particles being lighter than $\sim 1\ {\rm
TeV}$.\footnote
{In the focus point scenario of the supergravity model
\cite{PRD61-095004,PRL84-2322}, however, the lightest neutralino
becomes higgsino-like \cite{PLB482-388}.}
In fact, one sees that the lightest and the second lightest
neutralinos and the lighter chargino are quite degenerate in mass,
indicating that they are higgsino-like.

In the minimal anomaly mediation, the sign of the gluino mass is
opposite to that of the Wino mass. This causes a potential conflict
between the muon anomalous magnetic moment $a_{\mu}$ and Br$(b
\rightarrow s \gamma)$ \cite{PRD61-095004}.  The result reported by
the Brookhaven E821 experiment \cite{PRL86-2227} suggests a deviation
from the standard model prediction, indicating a new contribution from
new physics. If this is real, the new physics should give a positive
contribution to $a_{\mu}$, which constrains the sign of $\mu_HM_2 $
$(>0)$ in the context of SUSY. This choice also fixes the sign of
$\mu_HM_3 $ ($<0$ in the minimal anomaly mediation). This sign plays
an important role in the SUSY contributions to $b\rightarrow s\gamma$.
With this choice, the charged Higgs loop and the chargino/stop loop
give the contributions with the same sign, conflicting with the
experimental constraints. One of the advantages of our scenario is
that the gluino mass has the same sign as the Wino mass as far as
$N_5>3$, and thus the two constraints can be simultaneously satisfied.

Figs.\ \ref{fig:mdm12} and \ref{fig:mdm14} depict the SUSY
contribution to $a_\mu$ in the model 1 for $M_{\rm mess}=10^{12}$ and
$10^{14}\ {\rm GeV}$.  (For the formula for the SUSY contribution to
$a_\mu$, see \cite{SUSYg-2}, and for recent works see \cite{recentMDM}.)  In general, for a fixed value of of the
gaugino mass, the stau mass decreases as $\tan\beta$ increases.
Furthermore, as mentioned before, slepton mass relative to the gaugino
mass becomes smaller as we adopt smaller value of the messenger scale.
As a result, for relatively small $M_{\rm mess}$, we obtain an upper
bound on $\tan\beta$ to evade the experimental upper bound on the stau
mass, unless the overall scale of the SUSY breaking is large enough.
(End points of the lines in Fig.\ \ref{fig:mdm12} correspond to such
points.)  One consequence of this fact is that, when the messenger
scale is low, the SUSY contribution to the muon anomalous magnetic
moment is constrained to be relatively small, in particular, compared
to the deviation observed by the E821 experiment.  Once the messenger
scale is pushed up, however, the stau mass can be large enough and one
finds that the requisite contribution of $O(10^{-9})$ is easily
realized.  Notice that, for $M_{\rm mess}=10^{14}\ {\rm GeV}$,
$f_a=(10^{14}-10^{15})\ {\rm GeV}$ is required since the parameter
$\lambda$ has to be smaller than $O(0.1)$ in our model. (See Eq.\ 
(\ref{M_mess}).)  For such a large value of $f_a$, energy density of
the axion field becomes larger than the critical density if the
standard evolution of the Universe is assumed \cite{axionOmega}.  For
a possible cosmological scenario in this case, see the next
subsection.

We also studied the superparticle mass spectrum for the model 2. A
typical mass spectrum in this case is given in Table 2 as well. The
lightest superparticle in the MSSM sector is again the stau. Because
the axino is heavy in the model 2, the stau becomes the LSP in the
whole theory. Since such a charged LSP is ruled out cosmologically
\cite{KudoYama}, it must decay through $R$-parity
violation. Furthermore the superparticle masses are somewhat spread
compared to the previous case, which may cause the naturalness
problem. Thus we conclude that the model 2 is less attractive than the
model 1.
\subsection{Other Issues}
Let us return to the model 1, and consider the effects of the stau
decay into the axino. The axion-multiplet coupling to the matter is
suppressed by the decay constant $f_{a}$. Since we are considering
the hadronic axion, the axion coupling to leptons vanishes at tree
level. In fact, it arises at two-loop order, which originates from the
anomalous coupling of the axion to two photons. Equivalently one can
compute this coupling by using the superfield techniques. The axion
coupling is then contained in the field dependent wave function
renormalization of the lepton multiplet.  The same technique enables us
to compute the axino-lepton-slepton coupling in a simple matter as in 
Ref.~\cite{RSW}. Their
result is given for the right-handed slepton
\begin{eqnarray}
    {\cal L}_{\tilde{a} l \tilde{l}} 
    = C_l \frac{F_{\Phi}}{\langle X\rangle}
    \tilde{a} l \tilde{l}^{\dagger}
    + {\rm h.c.},
\end{eqnarray}
where
\begin{eqnarray}
    C_l =
    \frac{1}{8 \pi^2}\frac{2N_{\bf 5} (N_{\bf 5}+33/5)}{11} 
    \left[ \alpha_1^2(M_{\rm mess}) - \alpha_1^2 (m_Z) \right].
\end{eqnarray}
It follows from this that  the lifetime of the stau is approximately 
\begin{equation}
    \tau_{\tilde \tau} = N_{\bf 5}^2
    \left( \langle X\rangle/10^{13} \mbox{GeV} \right)^2
    \left(  200 \mbox{GeV}/ m_{\tilde \tau} \right)^3
    \mbox{sec}.
\end{equation}
Thus in collider experiments, the staus do not decay inside detectors,
and they are practically regarded as stable particles. Thus highly
ionized tracks produced by the heavy charged particles will be a
signature of this scenario \cite{PRD58-035001}.

Cosmological implications of our model are also interesting since
there exist various exotic particles in our model, like axion, axino,
and saxino, which may affect evolution of the Universe.  In
particular, in the model 1, saxino may significantly affect the cosmology,
contrary to the conventional scenario.  Therefore, let us comment on
this point.

In the model 1, there are two relatively light scalar fields, which
are axion and saxino.  The initial amplitude of these fields are in
general displaced from the minimum of the potential and they start to
oscillate as the Universe expands.  As a result, we should worry about
cosmological difficulties possibly caused by these coherent
oscillations.  The effects of the axion field have been extensively
studied.

The effects of the saxino is quite different from those of the axion,
since in our model the saxino is much heavier than the axion.
Consequently, the lifetime of the saxino becomes much shorter than
that of axion, and hence the saxino may decay before the present
epoch.  The saxino $\sigma$ dominantly decays into the axion pair, and
the lifetime is calculated as
\begin{eqnarray}
    \tau_{\sigma} = 
    \left[ \frac{1}{64\pi}\frac{m_X^3}{\langle X\rangle^2} 
    \right]^{-1} 
    \simeq 10\ {\rm sec}
    \times c_2^{-1} \lambda^{-1}
    \left( \frac{F_\phi}{\rm 100\ TeV} \right)^{-3}
    \left(\frac{N_{\bf 5}}{8}\right)^{-1/2},
\end{eqnarray}
where we approximated that the mass of the saxino is comparable to
$m_X$.  The lifetime is much longer than 1 sec as far as $\lambda\ll
1$ and $F_\Phi\lesssim 100\ {\rm TeV}$.  As a result, if the coherent
oscillation exists in the early Universe, it decays after the big-bang
nucleosynthesis (BBN) starts.  Therefore, if the energy density of
$\sigma$ is too large at the time of the BBN, it spoils the great
success of it.  Assuming that the initial amplitude of the saxino
field as $\sim f_a$, the saxino energy density normalized by the
entropy density is estimated as
\begin{eqnarray}
    \frac{\rho_{\sigma}}{s} \sim \frac{m_X^{1/2} f_a^2}{M_*^{3/2}}
    \sim 1\times 10^{6} \ {\rm GeV}
    \times \lambda^{5/2}
    \left( \frac{F_\Phi}{\rm 100\ TeV} \right)^{1/2}
    \left( \frac{N_{\bf 5}}{8} \right)^{-3/2}.
    \label{rho/s}
\end{eqnarray}
Notice that this ratio is independent of time.

Important constraint on the saxino abundance is from the
overproduction of $^4$He.  If the saxino energy density is too large
at the time of the neutron decoupling, it boosts up the expansion rate
of the Universe and increases the freeze-out temperature of the
neutron.  If this happens, the neutron number density after the freeze
out becomes larger, resulting in an overproduction of $^4$He.  To
avoid this problem, the ratio $\rho_{\sigma}/s$ should be much smaller
than the ratio $\rho_{\rm rad}/s$ at the time of the neutron
decoupling, and we obtain an upper bound on $\lambda$ of $\sim
O(10^{-4})$, corresponding to $M_{\rm mess}\lesssim 10^{9}\ {\rm
GeV}$.  As seen in Fig.\ \ref{fig:map}, with the messenger scale lower
than $\sim 10^{9}\ {\rm GeV}$, a large number of the messenger
multiplet should be introduced in order to avoid the tachyonic
sfermion mass, which makes the gauge coupling constants
non-perturbative below the GUT scale.  One might think that we may
have a consistent scenario of cosmology if $M_{\rm mess}\sim 10^{9}\ 
{\rm GeV}$.  In this case, however, the lifetime of the saxino is as
long as $\sim 10^5\ {\rm sec}$ and hence the saxino decays after
dominating the Universe.  Since the saxino dominantly decays into the
axions, energy density of the axion becomes 2 $-$ 3 orders of
magnitude larger than that of the radiation and hence there exists
large extra energy density in the form of the relativistic matter
(i.e., the axion).  First, this fact changes the epoch of the
radiation-matter equality to $z\sim 10$.  In addition, the axion is a
very weakly interacting particle and hence the cosmic density
fluctuation for the scale which enters the horizon before the equality
is washed out by the effect of the free-streaming.  These effects
cause a serious difficulty in the galaxy formation.  Furthermore, some
part of the saxino will decay into the pion pair through the QCD
anomaly.  Such pions dissociate the light elements produced by the BBN
and spoil the great success of the standard BBN.  Thus, we should
conclude that the scenario with $M_{\rm mess}\sim 10^{9}\ {\rm GeV}$
does not work.  Therefore, if we assume the conventional scenario of
cosmology, our model has a serious cosmological disaster.

This problem is, however, easily solved if a large amount of entropy
is produced after the onset of the saxino oscillation.  With a large
entropy production, the energy density of the saxino field is diluted
and the ratio $\rho_{\sigma}/s$ becomes much smaller than the value
given in Eq.\ (\ref{rho/s}).  In the anomaly-mediated models, the
moduli fields may be heavy and they can become a natural source of the
late-time entropy production.  It is also important to point out that,
if a large late-time entropy production occurs, axinos are also
diluted.  Thus, the axino relic density becomes much smaller than the
present critical density even if the axinos are thermally produced in
the early Universe.  In addition, if the entropy production occurs
after the QCD phase transition, coherent oscillation of the axion
field is also diluted and the cosmological bound on the axion scale
$f_a$ can be relaxed \cite{PLB383-313}.  In this case, $f_a$ much
larger than $10^{13}\ {\rm GeV}$ becomes possible.  This fact may be
of some help if a relatively large value of the SUSY contribution to
the muon magnetic moment is required, as mentioned in the previous
subsection.  It should be also noted that the relic staus are also
diluted by the entropy production.  Thus, the relic staus do not
significantly affect the BBN even if their lifetime is longer than
$\sim 1\ {\rm sec}$.  The scenario with the late-time entropy
production by the modulus decay may be tested by precisely measuring
the cosmic microwave anisotropy since the decay of the modulus field
may generate correlated mixture of the adiabatic and isocurvature
fluctuations \cite{hph0110096}.

In the model 2, the axino and saxino are as heavy as $\sim
O(F_\Phi)$. As a result, they are short-lived, and irrelevant in
cosmology.

\section{Summary}
\label{sec:summary}
\setcounter{equation}{0}

In this paper, we have considered hadronic axion models in the
supersymmetric theory where SUSY breaking is mediated by the
super-Weyl anomaly. SUSY breaking plays an important role in
determining the VEV of the axion multiplet $X$ which spontaneously
breaks $U(1)_{\rm PQ}$ symmetry. We constructed three models in each
of which the role of the SUSY breaking is different. The first model
has no superpotential for $X$ and thus the scalar potential solely
comes from SUSY breaking. In fact, the VEV of $X$ is determined by the
balance between the two terms: one is the negative soft SUSY breaking
mass generated by the super-Weyl anomaly, and the other is from the
higher order term in the K\"{a}hler potential.  In the second model,
the PQ symmetry would not be broken in the absence of the SUSY
breaking.  Once the SUSY breaking is switched on, the potential
minimum shift to the vacuum with the symmetry breaking. In the third
model, the PQ symmetry is already broken in the exact SUSY limit, but
the vacuum is degenerate along a non-compact flat direction. It is
then the effect of the SUSY breaking that lifts the flat direction,
thus fixing the PQ scale. In all three models, the PQ scale can be
adjusted to the allowed range by appropriate choice of the coupling
constants.

In the first two models, the PQ multiplet possesses non-trivial SUSY
breaking, and thus it mediates SUSY breaking to the MSSM sector via a
mechanism similar to the gauge mediation. Then soft SUSY breaking
masses are deflected from those from the pure anomaly mediation. Our
models provide a realization of the deflected anomaly mediation
advocated in Ref.\ \cite{JHEP9905-013}. On the other hand, in the
model 3 the $X$ field has trivial SUSY breaking, and thus it does not
affect the mass spectrum of the superparticles in the MSSM. Therefore
the sleptons remain tachyonic unless some other mechanism lifts up the
slepton masses.

Phenomenological issues of the models 1 and 2 are briefly
discussed. In these models, the stau becomes the lightest
superparticle in the MSSM sector.  In collider experiments, charged
tracks due to the stau will be an important signature of the models.
We also computed the SUSY contribution to the muon anomalous magnetic
moment. We found that, despite relatively heavy Winos, the
contribution from the SUSY sector can be comparable to or even larger
than that from the weak interaction sector. Thus it can easily explain
the possible discrepancy from the standard model prediction reported
by the E821 experiment.

One of the crucial differences between the two models is the mass of
the axino.  In the model 1, the axino is light and becomes the
lightest superparticle of the whole theory. In this case, the stau is
no longer stable, but decays to the axino, with lifetime typically
larger than 1 sec. In the model 2, the axino acquires mass at the tree
level and thus it is as heavy as the gravitino. In this case, the stau
will be the LSP of the whole sector and so we need $R$-parity
violation to avoid the charged stable particle.

We also discussed cosmology of the models. In model 1, the saxino is
light and long-lived.  We argued that its coherent oscillation would
spoil the success of the BBN. To avoid it, we invoke the late time
entropy production, and thus cosmological evolution of the model 1
should differ from the standard thermal history of the Universe. On
the other hand, in model 2, the saxino has a mass comparable to
$F_{\Phi}$, and thus is heavy. Thus it is short-lived, and
cosmologically harmless.

\section*{Acknowledgment}

The authors thank T. Takahashi for useful discussion.  This work was
supported in part by the Grant-in-aid from the Ministry of Education,
Culture, Sports, Science and Technology, Japan, priority area (\#707)
``Supersymmetry and unified theory of elementary particles,'' and in
part by the Grants-in-aid No.11640246 and No.12047201.


\begin{thebibliography}{99}

\bibitem{theta}
    V. Baluni,
    {\sl Phys.\ Rev.} {\bf D19} (1979) 2227.
    
\bibitem{PQ}
    R.D. Peccei and H.R. Quinn,
    {\sl Phys.\ Rev.\ Lett.} {\bf 38} (1977) 1440;
    {\sl Phys.\ Rev.} {\bf D16} (1977) 1791.
    
\bibitem{axion}
    S. Weinberg,
    {\sl Phys.\ Rev.\ Lett.} {\bf 40} (1978) 223;
    F. Wilczek,
    {\sl Phys.\ Rev.\ Lett.} {\bf 40} (1978) 279.
    
\bibitem{hadaxion}
    J.E. Kim, {\sl Phys.\ Rev.\ Lett.} {\bf 43} (1979) 103;
    M.A. Shifman, A.I. Vainshtein and V.I. Zakharov,
    {\sl Nucl.\ Phys.} {\bf B166} (1980) 493.

\bibitem{invaxion}
    M. Dine, W. Fischler and M. Srednicki, 
    {\sl Phys.\ Lett.} {\bf B104} (1981) 199;
    A.P. Zhitnitskii, Sov. J. {\sl Nucl.\ Phys.} {\bf 31} (1980) 260.

\bibitem{SUSYaxion}
    K. Rajagopal, M.S. Turner and F. Wilczek,
    {\sl Nucl.\ Phys.} {\bf B358} (1991) 447;
    T. Goto and M. Yamaguchi, 
    {\sl Phys.\ Lett.} {\bf B276} (1992) 103;
    E.J. Chun, J.E. Kim and H.P. Nilles, 
    {\sl Phys.\ Lett.} {\bf B287} (1992) 123;
    E.J. Chun and A. Lukas, 
    {\sl Phys.\ Lett.} {\bf B357} (1995) 43;
    T. Asaka and M. Yamaguchi,
    {\sl Phys.\ Lett.} {\bf B437} (1998) 51;
    {\sl Phys.\ Rev.} {\bf D59} (1999) 125003.

\bibitem{AMSB}
    L. Randall and R. Sundrum,
    {\sl Nucl.\ Phys.} {\bf B557} (1999) 79;
    G.F. Giudice, M.A. Luty, H. Murayama and R. Rattazzi,
    {\sl JHEP} {\bf 9812} (1998) 027;
    J.A. Bagger, T. Moroi and E. Poppitz,
    {\sl JHEP} {\bf 0004} (2000) 009.
    
\bibitem{Cosmo}
    T. Gherghetta, G.F. Giudice and J.D. Wells,
    {\sl Nucl.\ Phys.} {\bf B559} (1999) 27.

\bibitem{NPB570-455}
    T. Moroi and L. Randall,
    {\sl Nucl.\ Phys.} {\bf B570} (2000) 455.

\bibitem{JHEP9905-013}
    A. Pomarol and R. Rattazzi,
    {\sl JHEP} {\bf 9905} (1999) 013.

\bibitem{inverted}
    E. Witten,
    {\sl Phys.\ Lett.} {\bf B105} (1981) 267.

\bibitem{sequestered}
    L. Randall and R. Sundrum, in Ref.\ \cite{AMSB};
    M.A. Luty and R. Sundrum;
    {\sl Phys.\ Rev.} {\bf D62} (2000) 035008.

\bibitem{WB}
    See, for example, J. Wess and J. Bagger,
    {\sl Supersymmetry and supergravity}
    (Princeton University Press, 1992).

\bibitem{RSW}
R. Rattazzi, A. Strumia and J.D. Wells,
{\sl Nucl.\ Phys.}\ {\bf B576} (2000) 3.

\bibitem{PRD61-095004}
    J.L. Feng and T. Moroi,
    {\sl Phys.\ Rev.}\ {\bf D61} (2000) 095004.

\bibitem{PRL84-2322}
    J.L. Feng, K.T. Matchev and T. Moroi,
    {\sl Phys.\ Rev.\ Lett.} {\bf 84} (2000) 2322.

\bibitem{PLB482-388}
    J.L. Feng, K.T. Matchev and F. Wilczek,
    {\sl Phys.\ Lett.} {\bf B482} (2000) 388.

\bibitem{PRL86-2227}
    H.N. Brown et al.\ (Muon $g-2$ Collaboration), 
    {\sl Phys.\ Rev.\ Lett.}\ {\bf 86} (2001) 2227.

\bibitem{SUSYg-2}
    J.L. Lopez, D.V. Nanopoulos and X. Wang, 
    {\sl Phys.\ Rev.}\ {\bf D49} (1994) 366;
    U. Chattopadhyay and P. Nath,
    {\sl Phys.\ Rev.}\ {\bf D53} (1996) 1648;
    T. Moroi, 
    {\sl Phys.\ Rev.}\ {\bf D53} (1996) 6565; 
    Erratum-ibid, {\bf D56} (1997) 4424.

\bibitem{recentMDM}
    J.L. Feng and K.T. Matchev,
    {\sl Phys.\ Rev.\ Lett.}\ {\bf 86} (2001) 3480;
    L.L. Everett, G.L. Kane, S. Rigolin and L.-T. Wang,
    {\sl Phys.\ Rev.\ Lett.}\ {\bf 86} (2001) 3484;
    E.A. Baltz and P. Gondolo,
    {\sl Phys.\ Rev.\ Lett.}\ {\bf 86} (2001) 5004;
    U. Chattopadhyay and P. Nath,
    {\sl Phys.\ Rev.\ Lett.}\ {\bf 86} (2001) 5854;
    S. Komine, T. Moroi and M. Yamaguchi,
    {\sl Phys.\ Lett.}\ {\bf B506} (2001) 93;
    {\sl Phys.\ Lett.}\ {\bf B507} (2001) 224;
    J. Hisano and K. Tobe,
    {\sl Phys.\ Lett.}\ {\bf B510} (2001) 197;
    J.R. Ellis. D.V. Nanopoulos and K.A. Olive,
    {\sl Phys.\ Lett.}\ {\bf B508} (2001) 65;
    R. Arnowitt, B. Dutta, B. Hu and Y. Santoso,
    {\sl Phys.\ Lett.}\ {\bf B505} (2001) 177;
    K. Choi, K. Hwang, S.K. Kang, K.Y. Lee and W.Y. Song,
    {\sl Phys.\ Rev.}\ {\bf D64} (2001) 055001;
    S.P. Martin and J.D. Wells, 
    {\sl Phys.\ Rev.}\ {\bf D64} (2001) 035003;
    H. Baer, C. Balazs, J. Ferrandis and X. Tata,
    {\sl Phys.\ Rev.}\ {\bf D64} (2001) 035004;
    S. Baek, T. Goto, Y. Okada and K. Okumura,
    {\sl Phys.\ Rev.}\ {\bf D64} (2001) 095001;
    C.H. Chen and C.Q. Geng,
    {\sl Phys.\ Lett.}\ {\bf B511} (2001) 77;
    Z. Chacko and G.D. Kribs,
    {\sl Phys.\ Rev.}\ {\bf D64} (2001) 075015;
    T. Blazek and S.F. King,
    {\sl Phys.\ Lett.}\ {\bf B518} (2001) 109;
    G.-C. Cho and K. Hagiwara,
    {\sl Phys.\ Lett.}\ {\bf B514} (2001) 123;
    M. Byrne, C. Kolda and J.E. Lennon, hep-ph/0108122;
    S. Komine and  M. Yamaguchi, hep-ph/0110032;
    M. Endo and T. Moroi, hep-ph/0110383.

\bibitem{axionOmega}
    J. Preskill, M. Wise and F. Wilczek,
    {\sl Phys.\ Lett.}\ {\bf B180} (1983) 127;
    L. Abbott and P. Sikivie,
    {\sl Phys.\ Lett.}\ {\bf B180} (1983) 133;
    M. Dine and W. Fischer,
    {\sl Phys.\ Lett.}\ {\bf B180} (1983) 137.

\bibitem{KudoYama}
A.~Kudo and M.~Yamaguchi,
{\sl Phys.\ Lett.}\ {\bf B516} (2001) 151.

\bibitem{PRD58-035001}
    M. Drees and X. Tata,
    {\sl Phys.\ Lett.}\ {\bf B252} (1990) 695;
    J.L. Feng and T. Moroi,
    {\sl Phys.\ Rev.}\ {\bf D58} (1998) 035001.

\bibitem{PLB383-313}
    M. Kawasaki, T. Moroi and T. Yanagida,
    {\sl Phys.\ Lett.}\ {\bf B383} (1996) 313.

\bibitem{hph0110096}
    T. Moroi and T. Takahashi, hep-ph/0110096. 

\end{thebibliography}
\end{document}